\documentclass[11pt,a4paper]{article}

\usepackage{jcappub}
\usepackage{graphicx}
\usepackage{amssymb}
\usepackage{amsmath}
\usepackage{epsfig,multicol,bbm}
\usepackage{amsmath,amsfonts,amssymb,longtable,mathtools}
\usepackage{array}
\usepackage{graphicx}
\usepackage{subfig}
\usepackage{enumerate}
\usepackage{verbatim}
\usepackage{bm}
\usepackage{latexsym}
\usepackage{epsfig}
\usepackage{psfrag}
\usepackage{ulem}
\usepackage{textcomp}
\usepackage{color}
\usepackage{colortbl} 
\usepackage{placeins}
\usepackage{longtable,booktabs,caption}
\usepackage{changepage}
\usepackage{float}
\usepackage[utf8]{inputenc}

\makeatletter
\gdef\@fpheader{}
\makeatother

\def\be{\begin{equation}}
\def\ee{\end{equation}} 
\def\bea{\begin{eqnarray}}
\def\eea{\end{eqnarray}} 
\def\bseq{\begin{subequations}}
\def\eseq{\end{subequations}}
\def\ba{\begin{align}}
\def\ea{\end{align}}

\def\Mp{M_{_{\rm Pl}}}

\def\a1{\alpha_{_1}}
\def\b1{\beta_{_1}}
\def\de1{\delta_{_1}}
\def\g1{\gamma_{_1}}

\newcommand{\ie}{\textit{i.e.~}}

\begin{document}
\title{\huge Dark energy, D-branes, and  Pulsar Timing Arrays}
\author[a]
{Debika Chowdhury,}
\author[a,b]
{Gianmassimo Tasinato,}
\author[a]
{Ivonne Zavala}
\affiliation[a]
{Department of Physics, Swansea University, Swansea, SA2 8PP, U.K.}
\affiliation[b]
{Dipartimento di Fisica e Astronomia, Universit\`a di Bologna,  Italia}\emailAdd{debika.chowdhury@swansea.ac.uk} 
\emailAdd{g.tasinato@swansea.ac.uk}
\emailAdd{e.i.zavalacarrasco@swansea.ac.uk}
\abstract{Several pulsar timing array (PTA)
 collaborations  recently announced the first
 detection of a stochastic
gravitational wave (GW) background, leaving open the question of its source. 
We explore  the possibility that  it originates from cosmic inflation, a guaranteed
source of primordial GW. The inflationary GW background
 amplitude is enhanced at PTA scales by a non-standard
 early cosmological evolution,  driven by Dirac-Born-Infeld (DBI) scalar 
 dynamics motivated by string theory.    The resulting 
 GW energy density has a broken power-law frequency profile,  entering
 the PTA band  with
  a peak amplitude consistent with the recent GW detection.
  After this initial DBI kination epoch, the dynamics starts  a new phase 
  mainly  controlled by the scalar potential. It provides a realization of an early dark energy scenario aimed
  at relaxing the $H_0$ tension, and a late dark energy model which explains
the   current cosmological acceleration with no need of a cosmological constant. 
  Hence
   our mechanism --  besides
  providing a possible explanation for the recent PTA results -- 
  connects them with testable properties of the physics of the dark universe. 
  }

\maketitle

\section{Introduction}

Cosmic inflation, well tested at large CMB scales,  is a guaranteed source of primordial tensor fluctuations
 \cite{Grishchuk:1974ny,Starobinsky:1979ty,Rubakov:1982df,Fabbri:1983us,Abbott:1984fp}.
 However, their amplitude is typically too small to be directly detected by gravitational wave (GW) experiments. In this work we explore a post-inflationary mechanism capable of enhancing
the size  of inflationary tensor fluctuations at frequencies detectable by pulsar timing arrays (PTA)~(for other examples of post-inflationary mechanisms, see~\cite{Haque:2021dha,Ashoorioon:2022raz,Guo:2023hyp}). 
  The GW energy density  parameter $\Omega_{\rm GW}$ depends on the early-time behavior of the cosmological scale factor, which can be influenced by the presence
  of stiff matter \cite{Giovannini:1998bp,Boyle:2007zx,Boyle:2005se},  or scalar fields coupled conformally \cite{Salati:2002md,Catena:2004ba,Gelmini:2013awa,Lahanas:2006hf,Pallis:2009ed,Dutta:2016htz} or disformally \cite{Dutta:2017fcn,Chowdhury:2022gdc} to the metric.  
Extending the results of \cite{Chowdhury:2022gdc}, 
 we show that an
  early epoch of scalar domination
  -- motivated by a quantum gravity approach to early-universe cosmology --
   induces a non-standard cosmological evolution, which  enhances the amplitude of  $\Omega_{\rm GW}$ at frequencies around $10^{-8}$ Hz.   Our framework  provides
    a possible contribution towards the  explanation of the  recent detection
    of a stochastic GW background, as announced  by several PTA collaborations: NANOGrav \cite{NANOGrav:2023gor}, PPTA \cite{Reardon:2023gzh}, EPTA  \cite{Antoniadis:2023ott}, 
    CPTA \cite{Xu:2023wog} (see also e.g. \cite{NANOGrav:2023hvm,Antoniadis:2023xlr} for    some of their own
    preliminary interpretations  in terms of early universe sources). At the same
    time, it  may ameliorate
 some well-known cosmological problems related with the Hubble tension (see \cite{DiValentino:2021izs,Schoneberg:2021qvd} for reviews), and the nature of dark energy. 

\smallskip

 We start with a theoretical  section \ref{sec_setup} developing our string-motivated set-up based on the dynamics of a D-brane in a higher-dimensional space-time.
  The D-brane motion through extra-dimensional angular directions   is described in terms of an axionic field,  coupled  to four-dimensional 
gravity, and to matter living on the brane through a disformal coupling \cite{Bekenstein:1992pj}. 
 The scalar-tensor action includes a kinetic term of Dirac-Born-Infeld (DBI) form, which is instrumental in controlling an initial phase of  kinetically-driven scalar evolution that
 enhances the GW spectrum. Moreover, the Lagrangian includes a potential term,  responsible for driving both  an early  and   a late dark energy dominated epoch~(for an example of a cascading dark energy scenario, see~\cite{Rezazadeh:2022lsf}).    

\smallskip

We continue with a phenomenological section \ref{sec_pheno},
   which  starts examining   how the initial scalar dynamics drives an initial epoch of coupled DBI
  kination. This phase affects the early evolution of the Hubble parameter,  enhancing
 the size of the inflationary SGWB spectrum, which acquires a 
 broken power-law profile  with a peak amplitude well within the sensitivity curves  of PTA experiments. Interestingly, the characteristic
 scale controlling the  DBI kination  is comparable with the scale of QCD transition. We compare the peak amplitude of our profile with recent data from NANOGrav collaboration \cite{NANOGrav:2023gor}, finding overall agreement.

After examining the  consequences of our set-up for GW physics,
 in section \ref{sec_elde} we  discuss how the 
  initial DBI kination phase is connected to a stage where the scalar dynamics is mainly controlled by a 
 string-motivated axion potential. The structure of the axion potential is determined by the isometries of
 the  extra-dimensional space, as well as non-perturbative effects. It
 can be sufficiently rich to first drive 
 a phase characterized by early dark energy  -- which can address the Hubble tension  -- followed by a late dark energy epoch,  which can explain the current acceleration of our universe. Interestingly, the corresponding axion decay constants 
 acquire sub-Planckian values. They might
  represent an acceptable set-up for building models
 of dark energy within string theory (see e.g. \cite{Cicoli:2023opf} for a review).

\section{Our set-up}
\label{sec_setup}

In this section we  discuss the effective action for a string-motivated
D-brane system. It can be described in terms of a scalar-tensor theory 
with interesting consequences for cosmology and the physics of gravitational
waves. 
  Our scenario
is motivated by  
  D-brane scalar-tensor theories as discussed in \cite{Koivisto:2013fta}: we consider a (stack of) D-brane(s)  moving along the angular direction of an internal warped  compactification. The axion field $\phi$ is associated with
 the angular 
  position of the brane through the extra-dimensional  space.  The system is described by a
 scalar-tensor theory characterized by disformal couplings to matter fields on the
 brane \cite{Bekenstein:1992pj} (see \cite{Sakstein:2014isa} for a review).
The reader interested in more general scenarios -- including  conformal couplings between the scalar
and the metric -- 
 can consult  \cite{Dutta:2016htz,Dutta:2017fcn,vandeBruck:2020fjo,Chowdhury:2022gdc}. 

The scalar-tensor action we consider is:
\noindent
\be
S_{\rm tot}\,=\,S_{\phi}+S_{\rm m},
\ee
where
\bea
\label{defsf}
S_\phi&=&\int d^4 x\,\sqrt{-g}\,\left[\frac{R}{2 \kappa^2}-
M^4\,\sqrt{1+\frac{(\partial \phi)^2}{M^4}}
+M^4-{ V}(\phi)\right]\,,
\\
\label{smat}
S_{\rm m}&=&-\int d^4 x\,\sqrt{-g}\,{\cal L}_{\rm m}(\tilde g_{\mu\nu})\,.
\eea
 $\kappa^2=\Mp^{-2}=8\,\pi\,G$, and $M$ is a mass scale entering in the scalar kinetic term, related to the (possibly warped) tension of the D-brane, as well as other
 properties as warping and fluxes. Matter fields contained 
  in action \eqref{smat} are disformally coupled to the metric $g_{\mu\nu}$  entering 
  in eq \eqref{defsf} via
\be
\tilde g_{\mu\nu}\,=\,g_{\mu\nu}+\frac{\partial_\mu \phi \,\partial_\nu \phi}{M^4}\,.
\ee
 We 
assume that matter  is a perfect fluid, characterized by pressure and energy density only  (see \cite{Koivisto:2013fta} for details). 

The scalar kinetic terms have the characteristic Dirac-Born-Infeld (DBI) form of D-brane actions \cite{Polchinski:1996na}.
 The scalar potential $V(\phi)$ in eq \eqref{defsf} contains various contributions, which
 are periodic functions   of  the size of the internal angular dimensions.
  We focus on the case of a   D-brane fixed at a specific  value of the radial component
  along the  (warped) extra dimensions. Its position is also fixed  within     four internal angular directions, leaving the object   free to
  move  along  only one angular direction in the warped compact space. The motion of the brane through this angular direction is controlled 
by the axion field $\phi$   appearing in the scalar-tensor actions \eqref{defsf}.  See
 e.g.   \cite{Kenton:2014gma,Chakraborty:2019dfh} for 
 a scenario where this idea is  explicitly explored to realise natural inflation with D3 and D5 branes in a warped resolved conifold geometry \cite{Candel,RCmetric,Kleb}.
 
 \smallskip
 Let us be a bit more explicit about the structure of the axion potential. 
  The D-brane scalar potential  entering eq \eqref{defsf} acquires the schematic form \cite{Baumann:2010sx,Kenton:2014gma}:%
\bea
V(\theta) = \bar V(\rho_0)   + \delta \left(\overline{\Phi}_-(\rho_0) +\Phi_h(\rho_0,\theta) \right)\,,%
\eea
 with $\theta$ the angular direction along which the D-brane moves, that -- upon canonical normalization -- will give rise to the scalar field $\phi$ appearing in action \eqref{defsf}.
 The quantity 
 $\delta$ is a constant parameter depending on the type of D-brane considered. $\rho$  is the radial coordinate  fixed at $\rho=\rho_0$. 
 $\bar V$ and $\overline{\Phi}_-$ represent contributions to the D-brane  potential  depending only on the fixed radial coordinate $\rho_0$. Instead, $\Phi_h$ depends 
 on the angular coordinate as well, and for a warped resolved conifold is   given by \cite{Kenton:2014gma}:
\bea
\Phi_h (\rho_0,\theta,\phi_0)= \sum_{l=0}^{\infty}\sum_{m=-l}^{m=l}\left[a_lH_l^A(\rho_0) + b_lH_l^B(\rho_0) \right] Y_{lm}(\theta,\phi_0) \,. \label{Phih1} 
\eea
The indexes $(l,m)$ denote  
quantum numbers associated to 
isometries of the internal  geometry. 
The specific form of $H_l^A, H_l^B$ is not too important, as they are functions of the fixed radial D-brane position $\rho_0$, only.
 Considering for definiteness the values $(l,m)=(0,0), (1,0), (2,0), (3,0)$, we get \footnote{In \cite{Kenton:2014gma,Chakraborty:2019dfh}, only the  solutions  for $(l,m)=(0,0), (1,0)$, were kept, so that $\Phi_h$ took the form $ \Phi_h = A_1(\rho_0) + A_2(\rho_0) \,\cos{\theta}$.}
  \be
  \Phi_h = A_1(\rho_0) + A_2(\rho_0) \,\cos{\theta} + A_3(\rho_0) \,\cos^2{\theta}+A_4(\rho_0) \,\cos^3{\theta}\,.
  \ee
For suitable choices  of  the coefficients, 
 the potential for the canonically normalized field $\phi$ can be arranged to take the form\footnote{
Note that the choice of $l$ at this stage is arbitrary. This choice may be realised for example if all coefficients $a_l,b_l$ vanish for $l>3$ for some reason. In order to determine whether these coefficients can indeed vanish, a rigorous embedding of the model in a string compactification will be required. }
 \be\label{Vede}
V_1(\phi) = V_{0_{\rm ede}} \left(1-\cos[\kappa\,\phi/f_1]\right)^3\,.
\ee
 The quantity
 $f_1$ is an axion decay constant: we express it in terms of a dimensionless number, pulling out a factor of $\kappa$ in eq \eqref{Vede}. This is  the structure  of the potential  recently introduced for the so called {\it early dark energy}\/ \cite{Poulin:2018dzj,Poulin:2018cxd} to relax the $H_0$ discrepancy. We will discuss this topic in more detail in section \ref{sec_elde}. 

Besides the term in  eq \eqref{Vede}, 
  there may be additional non-perturbative contributions to the effective potential,  originating from bulk physics, which  generate extra  terms periodic in $\phi$. We assume  that
  such  contributions are  present, and we include an additional  term to the total axion potential, as~\cite{Kaloper:2005aj, Planck:2015bue}:
\be\label{Vde}
V_2(\phi) = V_{0_{\rm de}} (1-\cos[\kappa \phi/f_2]) \,.
\ee  
with $f_2$ an extra dimensionless decay constant. 
To summarize, 
in what follows we consider the following total potential for the scalar field as a sum of
two independent contributions
\bea
V(\phi)&=&V_1(\phi)+V_2(\phi) \,,
\\
&
=& V_{0_{\rm ede}} \left(1-\cos[\kappa \phi/f_1]\right)^3 +V_{0_{\rm de}} \left(1-\cos[\kappa \phi/f_2]\right)\,.
\label{Vtot}
\eea
We plug the potential \eqref{Vtot} in the total action \eqref{defsf}, and  study the consequences of this system for the cosmological  evolution
of our universe.

\subsection*{Evolution equations}
The equations of motion for the scalar field and the scale factor in a Friedmann-Lemaitre-Robertson-Walker metric, as derived from the Einstein-frame action \eqref{defsf}, \eqref{smat}, are obtained to be
 \cite{Dutta:2016htz,Dutta:2017fcn,Chowdhury:2022gdc}:
\begin{subequations}
\begin{align}
H^2 & =\frac{\kappa^2}{3}\frac{\,(1+\lambda)}{B} \rho\,, \label{eq:Fried2}\\
H_{_N} &= -H\left[\frac{3\,B}{2(1+\lambda)}\left(1+{w} \right) + \frac{\varphi_{_N}^2}{2}\,\gamma\right], 
\label{eq:HN}\\
  \varphi_{_{NN}} & \left[1 +\frac{\gamma^{-1}}{M^4}\,\frac{3\,B\,H^2}{\kappa^2(1+\lambda)}\right] + 3\,\varphi_{_N}\left[\gamma^{-2}-\frac{{w}}{M^4\,\gamma}\,\frac{3\,B\,H^2}{\kappa^2(1+\lambda)}\right]  \nonumber\\
  &\hskip 3cm + \frac{H_{_N}}{H}\,\varphi_{_N}\left[1+\frac{\gamma^{-1}}{M^4}\,\frac{3\,B\,H^2}{\kappa^2(1+\lambda)}\right]  + \frac{3\,B\,\lambda}{\gamma^3\,(1+\lambda)}\,\frac{V_{,\varphi}}{V}= 0,
  \label{eq:phiN}
\end{align}
\end{subequations}
where the subindex $N$ indicates derivatives with respect to the number of e-folds $dN=Hdt$.
For convenience we 
make use of the dimensionless scalar quantity $\varphi\,\equiv\,\kappa\,\phi$.  We introduce the scalar-dependent Lorentz factor $\gamma$, the quantity that characterize  DBI models:
\begin{equation}\label{eq:gamma-Dbrane}
 \gamma^{-2} = 1 - \frac{H^2}{M^4\kappa^2}\,\varphi_{_N}^2, 
\end{equation}
as well as the combinations
\bea
\label{eq:B-Dbrane}
 B &=& 1-\frac{\gamma^2\,\varphi_{_N}^2}{3\,(\gamma+1)}\,,
 \\
\lambda &=& \frac{ V}{\rho}.
\eea
 The parameter $\lambda$
 is the characteristic quantity that controls the size of the axionic potential term with respect to
 the total energy density.

 Since we are interested in comparing the expansion rates between our modified cosmological evolution and standard
 $\Lambda$CDM cosmology, we work in 
  the Jordan frame where energy-momentum and entropy are conserved.  (See the discussion in \cite{Chowdhury:2022gdc}.) The energy density, pressure and equation of state
  in this frame read
 \be
     \tilde{\rho} = \gamma^{-1}\,\rho, \qquad
     \tilde{p} = \gamma\,P, \qquad 
     \tilde{w} = w\,\gamma^2, 
     \ee
where the non-tilded quantities are computed in the Einstein frame.  Moreover,  the quantity  $\tilde \rho = \sum_i\tilde \rho_i$, is the total background energy density: the index $i$ runs over matter and radiation, and $\tilde \omega$ takes into account of the degrees of freedom  at a given temperature during cosmic evolution (see e.g.~\cite{Chowdhury:2022gdc}).
The departure from the standard cosmological evolution can thus be parameterized by the ratio  of Hubble parameters in our scalar-tensor gravity, with respect to its value in General Relativity (GR) (i.e. in absence of scalar contributions):
\be\label{xi}
 \frac{\tilde H}{H_{GR}} =\frac{\gamma^{3/2}  (1+\lambda)^{1/2}}{B^{1/2}}\,.
\ee

The evolution equations \eqref{eq:Fried2}-\eqref{eq:phiN} contain contributions that modify  the evolution equation with respect to $\Lambda$CDM. Any early-universe modification from the standard cosmological  evolution should not occur during or after Big-Bang Nucleosynthesis (BBN), to avoid
spoiling  its successful predictions. Hence   we  require that any deviations  from  $\Lambda$CDM  terminate
before
 the onset of BBN: i.e.,  when the temperature of the universe reaches values around $ T \sim$ MeV. 
 We  also investigate, however, how our system leads to late scalar-tensor contribution closely mimicking the cosmological
 constant of $\Lambda$CDM, driving the current acceleration of our universe. 
The structure of evolution equations \eqref{eq:Fried2}-\eqref{eq:phiN} is such that any  departure from the
equation-of-state value $\tilde w =1/3$  does not give a noteworthy effect in the scalar  evolution \footnote{In contrast to the conformal case \cite{Damour:1993id,Catena:2004ba}.}. In fact, in our case the most important effects in the very early  evolution 
of the axion field $\phi$ -- when the potential term contributions are negligible -- are associated with the DBI form of its kinetic terms \footnote{The initial phase of DBI kination
is distinct from  kination models considered in the literature (see e.g.
 \cite{Ferreira:1997hj,Redmond:2018xty,Co:2021lkc,Gouttenoire:2021jhk}).}. During
this phase, the amplitude of primordial gravitational waves is enhanced.  
 As the size of the scalar potential $V(\phi)$ becomes important, additional interesting phenomenological consequences occur.
 These topics are the subjects of next section.

\section{Phenomenology of the scalar-tensor theory}
\label{sec_pheno}

The cosmological evolution of the axionic field $\phi$  has
several interesting phenomenological consequences. We assume that the initial conditions
for  $\phi$ are set at very high temperatures, some time after cosmic inflation. We select them such that the  scalar has an initial kinetic-driven dynamics, capable of enhancing the spectrum of gravitational waves -- see section \ref{sec_enh}. The DBI kination phase is smoothly followed by a potential-driven phase, during which the scalar dynamics source
phases of early and late dark energy domination -- see section \ref{sec_elde}.

\subsection{Enhancing the gravitational wave signal at PTA scales}
\label{sec_enh}

\begin{figure}
\begin{center}
\includegraphics[width=7.5cm]{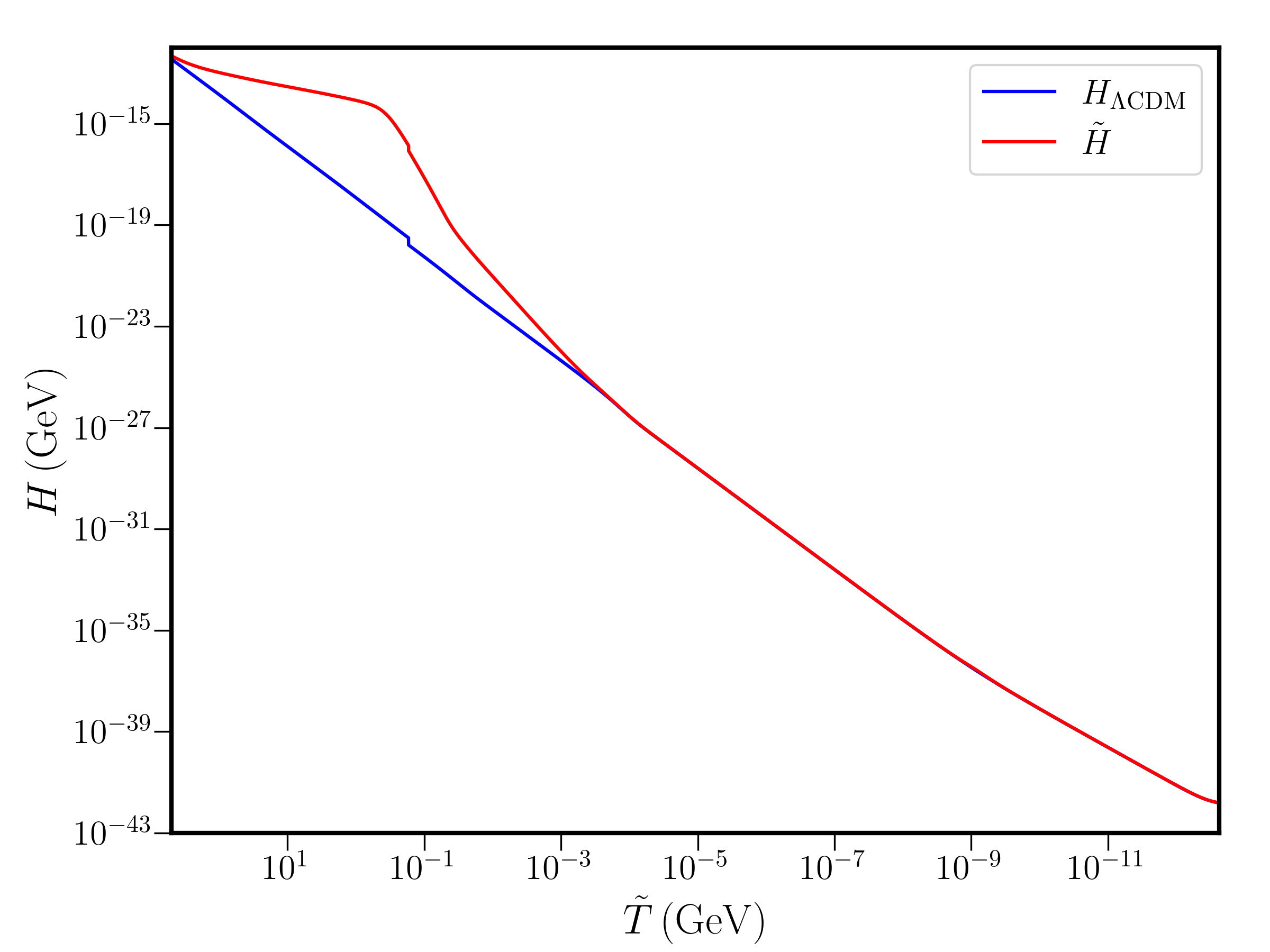}
\includegraphics[width=7.5cm]{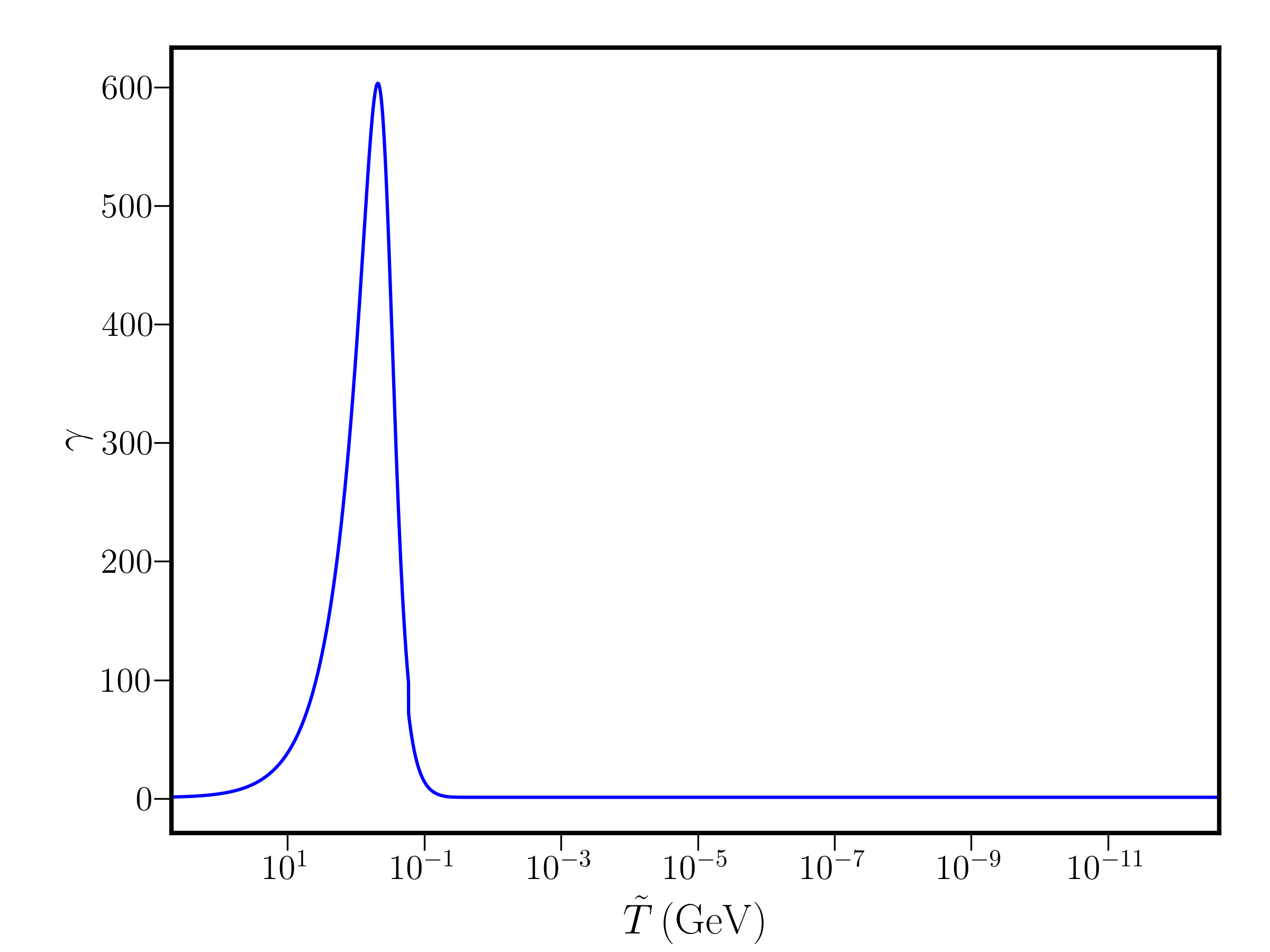}\\
\end{center}
\caption{\rm {\bf Left panel:} Evolution of the Hubble parameter in our scalar-tensor set-up (red line) and in $\Lambda$CDM case (blue  line), plotted as functions of temperature.
 {\bf Right panel:} Evolution of the Lorentz factor $\gamma$   starting from the initial conditions in Table~\ref{tab1}. }
\label{fig:Hgamma}
\end{figure}

 We show how an early modification
of standard evolution associated with the DBI-type action \eqref{defsf} amplifies the amplitude 
of inflationary gravitational
waves at PTA scales. We exploit the fact that this quantity is affected
by an early-time non-standard  cosmological evolution.  
The scalar-tensor dynamics, as controlled by the evolution equations of section
\ref{sec_setup}, starts soon after inflation ends. We assume that initial conditions
are chosen such that  initially the scalar  is driven by the DBI kinetic terms only, as appearing
in action \eqref{defsf}, with negligible
contribution from the potential terms (see e.g.  \cite{Dutta:2016htz,Dutta:2017fcn,Chowdhury:2022gdc}). 
The  parameter controlling the kinetic part of the action is  
  $M$, whose value can chosen together with the   initial conditions for $H_i$, $\varphi_i$, $\varphi_N^i$ -- the last quantity entering in the initial value  $\gamma_i$ for the DBI Lorentz parameter of eq \eqref{eq:gamma-Dbrane}. The initial value for $H$ is determined as described in Appendix \ref{App1}, starting from an initial value for the DBI parameter $\gamma_i$ of order ${\cal O}( 1)$. Once $\varphi_N^i$ is fixed along with an initial temperature $T_i$ -- which is associated with the initial  value of the scale factor    
   --  the value of $M$ is bounded from below by requiring that the solutions for $H^2$ be real and positive and such as not to spoil
 the predictions of BBN. 
Assuming that entropy is conserved, the relation between the temperature of the universe $T$ and the scale factor $a$ is
  \be
  \frac{a}{a_0}\,=\,\left( \frac{g_{* s,0}}{g_{* s}}\right)^{1/3}\,\frac{T_0}{T}\,,
  \ee
  where the index $0$ indicates quantities evaluated today. 
\begin{table}[h!]
\centering
 \begin{tabular}{| c | c | c | c | c |} 
 \hline 
 \cellcolor[gray]{0.7} $\varphi_i$ & \cellcolor[gray]{0.7} $\varphi_N^i$  &\cellcolor[gray]{0.7}  $H_i$ & \cellcolor[gray]{0.7} $T_i$ &\cellcolor[gray]{0.7}  $M$ \\  
 \hline\hline
 $0.2$ & $3.9\times 10^{-7}$ & $3.75359\times 10^{-13}$ GeV & $499.8043$ GeV & $765$ MeV \\
 \hline
   \end{tabular}
\caption{\small Initial conditions and disformal scale (recall that $\varphi$ is dimensionless and measured in Planck units).  }
\label{tab1}
\end{table}

We select initial conditions 
  as in Table \ref{tab1}.
 The initial conditions for $\varphi$ and the Hubble parameter are chosen such as to lead to an initial steady growth of the DBI Lorentz factor $\gamma$ of eq \eqref{eq:gamma-Dbrane}, and a transitory large deviation of the Hubble parameter from its  GR value. See Fig \ref{fig:Hgamma}. 
  We select  the parameter  $M$  demanding  
   that the scalar evolution does not
  interfere with BBN, which happens around 1 MeV -- see  Fig \ref{fig:Hgamma}.
  The value of $M$
   turns out to be of the  order of the QCD scale of $170$ MeV.
  Recall that we work in the Jordan frame -- see section \ref{sec_setup} -- hence
  with tilded quantities. 
  
  The initial enhancement of the Lorentz factor, as well as the early modifications of the Hubble parameter,  leads to an amplification of inflationary gravitational waves. The fractional energy density of primordial gravitational waves measured today is given by (we follow the treatments in \cite{Bernal:2020ywq,Watanabe:2006qe,Saikawa:2018rcs,Bernal:2019lpc}): 
  \bea
\tilde  \Omega^{0}_{\rm GW}(k)&\equiv&\frac{1}{\rho_c^0}\,\frac{d\,\tilde \rho_{\rm GW}^0(k)}{d \ln k}
  \\
  &\simeq&\frac{1}{24}\,{\cal P}_T(k)\,\left( \frac{ \tilde a_{\rm hc}}{\tilde a_0}\right)^4\,\left( \frac{\tilde H_{\rm hc}}{\tilde H_0}\right)^2 \label{newOGWf}
  \eea
  where  ${\cal P}_T$ is the primordial inflationary tensor spectrum, and the suffix `${\rm hc}$' indicates
  horizon crossing time for the mode $k$. The quantity  ${\cal P}_T$ is 
  \be
  {\cal P}_T(k)\,=\,\frac{2 \,H^2}{\pi^2\,M_{\rm Pl}^2} \Big|_{k=aH}\,,
  \ee
  and we take its amplitude at CMB scales to be ${\cal P}_T \,= \,  r A_S$, with $A_S = 2.1\times 10^{-9}$. For simplicity we assume that $r$ saturates the current upper bound $r=0.036$ provided  by the BICEP/Keck collaboration \cite{BICEP:2021xfz}. 
  
  Formula \eqref{newOGWf} indicates that any deviation of the cosmological evolution from standard $\Lambda$CDM
  can change the predictions for $\tilde \Omega^{0}_{\rm GW}(k)$,
  and possibly amplifies the spectrum of inflationary GW. In fact, we  make use of the evolution equations \eqref{eq:Fried2}-\eqref{eq:phiN}, and re-express  $\tilde \Omega^{0}_{\rm GW}$ as
  \be
  \label{secomgw}
  h^2\,\tilde \Omega^{0}_{\rm GW}\,=\,\left( \frac{{\cal P}_T }{24} \right)\,\left( \frac{ \tilde a}{\tilde a_0}\right)^4\,\frac{\gamma^3\,H_{\rm GR}^2}{B\,(H_0/h)^2}\,.
  \ee
  The quantity $\gamma$ is given in eq \eqref{eq:gamma-Dbrane}, while $B$  in eq \eqref{eq:B-Dbrane}.  $H_{\rm GR}$ corresponds to the GR Hubble parameter  in absence of scalar field contributions. Expression \eqref{secomgw}  shows
  that an enhancement of the DBI Lorentz factor $\gamma$, and a modification of the Hubble
  parameter with respect its GR value influence the scale-dependence of $ \Omega^{0}_{\rm GW}$. We can express
  this quantity as function of frequency $f=2 \pi \,k\,a_0$, through the formula
 \cite{Watanabe:2006qe,Saikawa:2018rcs,Kamionkowski:1993fg}
 \be
 f\,=\,2.41473\times 10^{23}\,\left(\frac{T_0}{T_{\rm hc}}\right)\,\left(\frac{g_{* s,0}}{g_{* s,{\rm hc}}}\right)^{1/3} \,\sqrt{\frac{8 \pi \rho_{\rm hc}}{3 M_{\rm Pl}^2}}\,{\rm Hz}\,,
 \ee
  where recall that {\rm hc} is the horizon crossing scale of the mode $k$.

\begin{figure}
\begin{center}
\includegraphics[width=10.5cm]{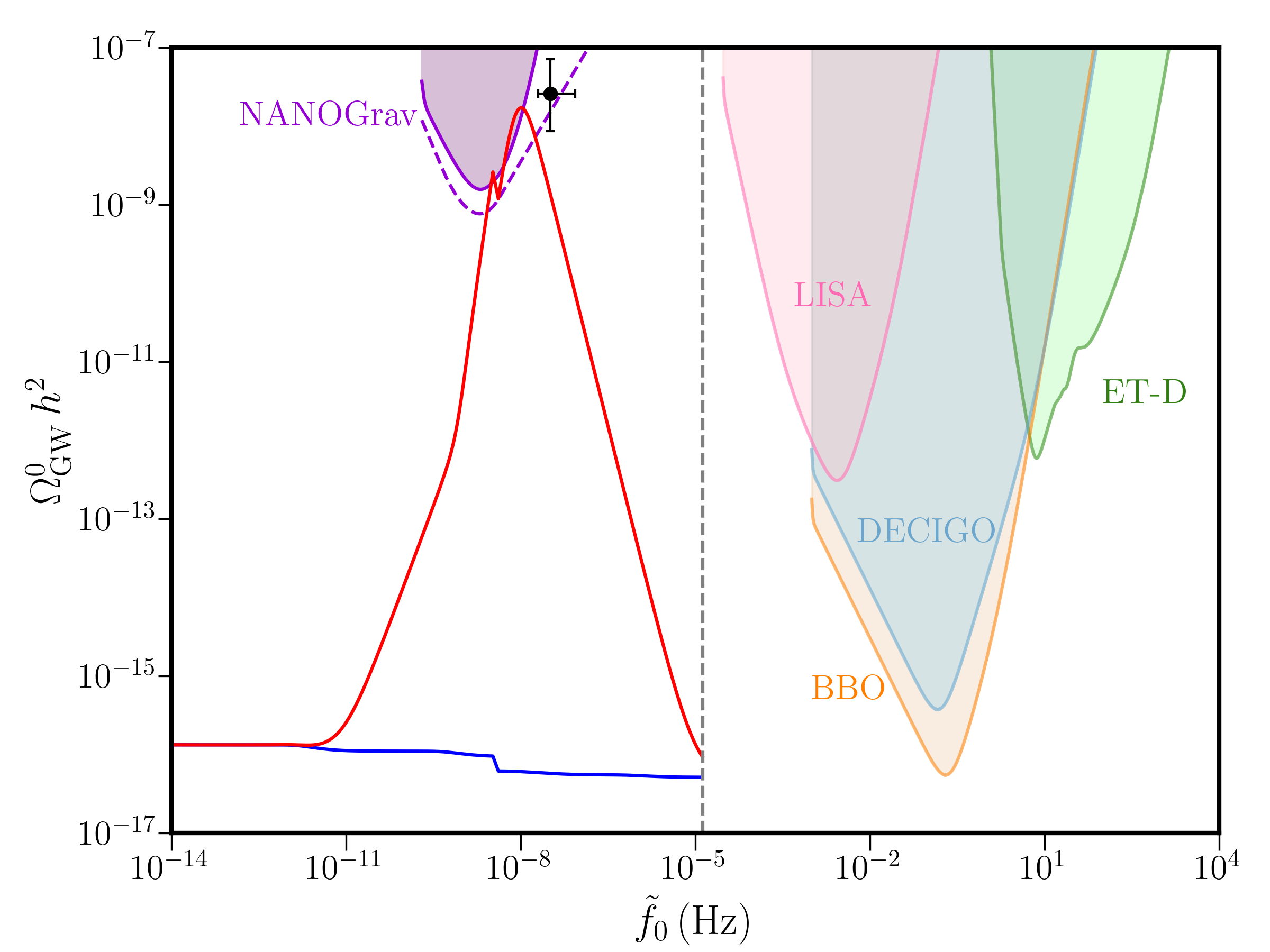}\\
\end{center}
\caption{\small
The profile of the primordial GW energy density of eq \eqref{secomgw} in our DBI-driven scalar-tensor
system (red line), as compared with the standard prediction in $\Lambda$CDM (blue line). 
  The dashed  vertical line indicates the
frequency corresponding to the initial temperature $T_i$ at which the scalar evolution starts. The  sensitivity curves for various experiments are 
  plotted using the notion of broken power-law sensitivity (BPLS) curve~\cite{Chowdhury:2022gdc} for ET-D, LISA, BBO, and DECIGO experiments with the condition of SNR $\geq 10$. For NANOGrav, the sensitivity curves have been constructed using the corresponding noise spectra for the stochastic gravitational wave background~\cite{NANOGrav:2023ctt,the_nanograv_collaboration_2023_8092346}. The light violet shaded portion corresponds to the BPLS sensitivity region with SNR $\geq 5$, while the dashed violet line denotes the boundary of the power-law sensitivity (PLS) region with SNR $\geq 5$. The black point along with the y-axis error bar denotes the value $\Omega_{_{\rm GW}}\,h^2=2.6^{+4.5}_{-1.7}\times 10^{-8}$ obtained with a model of the timing-residual power spectral density with variable power-law exponent, as described by the NANOGrav collaboration~\cite{NANOGrav:2023gor}. This value is derived from a fiducial power-law model, which explains why it lies outside the BPLS region, but within the allowed PLS region. The x-axis error bar denotes the $90\%$ confidence region of $3.2^{+5.4}_{-1.2}\times 10^{-8}$ Hz for the break frequency in a broken power-law model as obtained by the NANOGrav collaboration~\cite{NANOGrav:2023gor}. The peak of the GW spectrum reaches a maximum value of $\sim 1.7\times 10^{-8}$,   in agreement with the above bound, while the peak in the profile occurs at around $\sim 1.02\times 10^{-8}$ Hz.}
\label{fig:GW}
\end{figure}

We 
represent in Fig \ref{fig:GW} the GW spectrum obtained by numerically solving  the evolution equations of section \ref{sec_setup}, and plugging the results in eq \eqref{secomgw}.  The initial conditions in Table \ref{tab1} lead to a rapid, transient increase of $\gamma$, and allow us  to amplify the GW signal at PTA frequencies.  In fact,
the energy density associated with primordial gravitational waves
is raised by several orders of magnitude with respect to its standard value, 
for a frequency  around the $10^{-9} - 10^{-8}$ Hz band that is  probed by PTA experiments. 
The frequency profile of the spectrum acquires a broken power-law  shape. It initially
rises as $f^2$, to then grow\footnote{Note that the gravitational wave amplitude is enhanced purely due to the DBI kination period encoded in $\gamma, B$ in equation \eqref{secomgw}. Thus the enhancement behaviour is different from that of 
a standard ``kination" dominated epoch  \cite{Ferreira:1997hj,Redmond:2018xty,Co:2021lkc,Gouttenoire:2021jhk}. See  \cite{Chowdhury:2022gdc} for details.} as $f^5$ up to the peak, and then decreases as $f^{-3}$. 
The peak amplitude is of the same order as the value detected by the NANOGrav collaboration~\cite{NANOGrav:2023gor}. However, the NANOGrav value we compare with
is based on a fiducial power-law model, and a more sophisticated data
analysis would be needed for comparing our broken power-law shape with the amplitude
obtained by PTA
data. In fact,  \cite{NANOGrav:2023gor} also provides a brief analysis of broken power-law
models, providing the best-fit value for the break of the frequency profile: our result is
consistent with their value.

Figure \ref{fig:GW} also contains the sensitivity curves for the NANOGrav  experiment, 
as well as other detectors for reference. The sensitivity curves
are built with the broken power-law sensitivity (BPLS) curve technique, introduced in \cite{Chowdhury:2022gdc} as
an extension of the traditional power-law sensitivity curves of \cite{Thrane:2013oya}. (Our definition and
methods to obtain  BPLS is
slightly different from \cite{Schmitz:2020syl}.) The BPLS curve allows one to visually realise whether a broken power-law
signal can be detected by a given experiment: our profile for $\tilde \Omega_{\rm GW} h^2$ enters
into the sensitivity curve of NANOGrav, showing that the scalar-tensor theory we
described allows us to amplify the primordial spectrum of inflationary tensor fluctuations at a   level detectable by PTA experiments. We conclude that  our signal might contribute to the stochastic
GW background recently detected by PTA experiments \cite{NANOGrav:2023gor,Reardon:2023gzh,Antoniadis:2023ott,Xu:2023wog}. 

\begin{figure}
\begin{center}
\includegraphics[width=8.5cm]{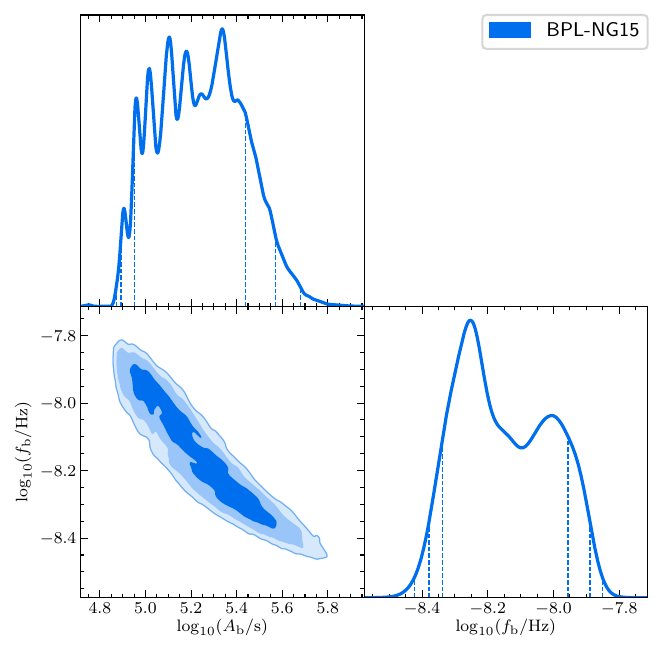}
\end{center}
\caption{The lower left panel in this plot shows the $68\%$, $95\%$, and $99\%$ confidence levels of the $2$-dimensional posterior distribution for the parameters of the broken power-law fitting function: ${\rm log}_{10}(A_{\rm b}/{\rm s})$ and ${\rm log}_{10}(f_{\rm b}/{\rm Hz})$. The other two panels show the $1$-dimensional marginalized posterior distributions for these parameters. The dashed vertical lines in the plots for the $1$-dimensional distributions indicate the $68\%$, $95\%$, and $99\%$ credible intervals. The uniform priors used in this analysis are: ${\rm log}_{10}(A_{\rm b}/{\rm s}) \in [2.0,8.0]$ and ${\rm log}_{10}(f_{\rm b}/{\rm Hz}) \in [-10.0,-6.0]$. We also assume Hellings \& Downs correlations between the pulsars.}
\label{fig:data-analysis}
\end{figure}

In order to compare the results of our D-brane disformal scenario with the recent NANOGrav 15-year data, we model the peak of the broken power-law GW spectrum as obtained in Fig.~\ref{fig:GW} as follows~\cite{NANOGrav:2020bcs,NANOGrav:2023gor}:
\begin{equation}
 \tilde{\Omega}_{\rm GW}^0\,h^2 = A_{\rm b}^2\,f^2\left(\frac{f}{f_{\rm yr}}\right)^{\sigma-3}\left[1+\left(\frac{f}{f_{\rm b}}\right)^{1/\varepsilon}\right]^{\varepsilon(\mu-\sigma)},
\end{equation}
where $f_{\rm yr}=1/{\rm year}$ in Hz.
On fitting this model to our numerical results, we find the following values of the parameters: ${\rm log}_{10}(A_{\rm b}/{\rm s})=5.279$, ${\rm log}_{10}(f_{\rm b}/{\rm Hz})=-8.048$, $\sigma=6.0$, $\varepsilon=0.25$, and $\mu=-2.0$. Keeping the spectral indices fixed, we vary ${\rm log}_{10}(A_{\rm b}/{\rm s})$ and ${\rm log}_{10}(f_{\rm b}/{\rm Hz})$ to arrive at the best-fit values for these quantities by using the publicly available code PTArcade~\cite{andrea_mitridate_2023_8106173,Mitridate:2023oar,Lamb:2023jls}. The results have been plotted in Fig.~\ref{fig:data-analysis}. According to these results, the Bayesian estimates for the best-fit values are: ${\rm log}_{10}(A_{\rm b}/{\rm s})=5.242 \pm 0.191$ and ${\rm log}_{10}(f_{\rm b}/{\rm Hz})=-8.149 \pm 0.144$. By comparing these limits with our aforementioned fitting parameters, we find that they are well in agreement with the NANOGrav data.

\subsection{Early and late dark energy}
\label{sec_elde}

Following the initial phase of DBI kination domination enhancing the GW 
spectrum (see Fig \ref{fig:GW}), the DBI Lorentz factor $\gamma$ returns
to small values (see Fig \ref{fig:Hgamma}). The subsequent scalar dynamics  is
smoothly matched to a second phase, 
 mainly controlled
by the potential terms of eq \eqref{Vede}. The potential-dominated
phase allows us to realize a scenario including both an early dark energy (EDE) 
 epoch, as well a late dark energy (LDE) phase which  explains the present day acceleration
 of the universe\footnote{Recently, a connection between EDE and stochastic GW signals
 has been explored in \cite{Kitajima:2023mxn}. Their framework differs from ours, since we
 use the properties of DBI kinetic terms for enhancing the GW spectrum -- see section \ref{sec_setup} -- while they make use of the properties of the EDE axionic potential. The peak
  of their GW signal is far from PTA scales.}.

The scalar potential \eqref{Vede} has been recently considered in \cite{Poulin:2018dzj,Poulin:2018cxd} as a possible way to relax the so called $H_0$ tension in cosmology, via the injection of an early period of dark energy domination, driven by the scalar $\phi$, called {\it early dark energy}\/ (EDE) (see \cite{Wetterich:2004pv,Doran:2006kp} for early EDE models). 
The $H_0$ tension refers to the discrepancy between local measurements of the Hubble parameter today from supernovae, and its inferred value from   cosmic
microwave background (CMB)  measurements assuming the standard $\Lambda$CDM  cosmological model (for recent reviews on the $H_0$ tension and solutions see \cite{DiValentino:2021izs,Schoneberg:2021qvd}. For recent reviews on EDE models see \cite{Kamionkowski:2022pkx,Poulin:2023lkg}). 
As discussed in section \ref{sec_setup}, in our case the potential also includes an additional term, leading to  a LDE domination driven by the  axion \cite{Kim:2002tq}. In total, the potential expressed in terms of the dimensionless scalar field $\varphi$ can be written as\footnote{Possible embeddings of EDE from string theory based on closed string moduli has been discussed recently in \cite{McDonough:2022pku,Cicoli:2023qri}.}
 (recall that the decay constants are dimensionless and given in Planck units):
\be
\label{secvP}
V(\varphi)= V_{0_{\rm ede}} \left(1-\cos[\varphi/f_1]\right)^3 +V_{0_{\rm de}} \left(1-\cos[\varphi/f_2]\right)\,.
\ee
The first term is responsible for the EDE epoch, and we
select $V_{0_{\rm ede}}\sim {\rm eV}^4$; the second term leads to late
time acceleration, hence $V_{0_{\rm de}} \sim (0.002\, {\rm eV})^4$. 
The axionic evolution controlled by the potential terms is preceded
by the DBI kinetic evolution of section \ref{sec_enh}: the requirement of matching the two
phases partially  determines the parameters of the model, as in Table  \ref{tab2}. 
Interestingly, this matching connects the early DBI kination phase which enhances
the GW signal, to a later evolution contributing to the physics of the dark universe. 

\begin{table}[h!]
\centering
 \begin{tabular}{| c | c | c | c  |} 
 \hline 
 \cellcolor[gray]{0.7} $V_{0_{\rm ede}}$ & \cellcolor[gray]{0.7} $V_{0_{\rm de}}$  &\cellcolor[gray]{0.7}  $f_1$ & \cellcolor[gray]{0.7} $f_2$  \\  
 \hline\hline
 $(6.32\times10^{-10}\,{\rm GeV})^4$ & $1.45247\times 10^{-47}\,{\rm GeV}^4$ & $0.3569 $ & $0.1044$     \\
 \hline
   \end{tabular}
\caption{\small The values of the parameters in the scalar potential \eqref{secvP}.}
\label{tab2}
\end{table}

\begin{figure}
\begin{center}
\includegraphics[width=9.0cm]{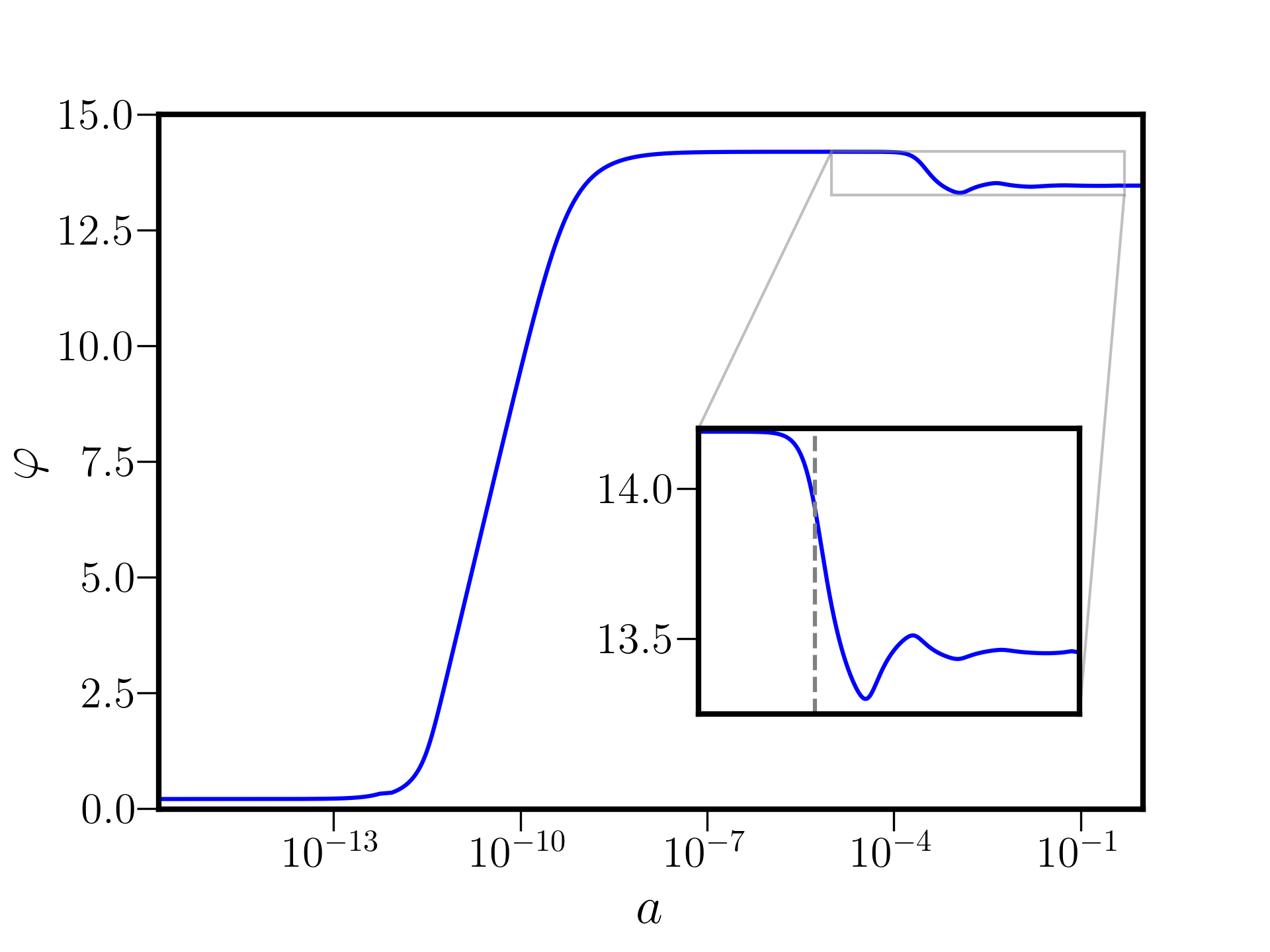}\\
\end{center}
\caption{\small The evolution of the scalar field $\varphi$  as a function of the scale factor in our set-up. The  initial conditions and parameter values can be found  in Tables~\ref{tab1} and ~\ref{tab2}. In the inset, we have zoomed into the region where the field oscillates around the minimum of the EDE potential. The dashed vertical line in the inset corresponds to a redshift of $z\sim 10^{3.562}$, which is around the time when the fractional energy density of the EDE component reaches its maximum value (see Fig.~\ref{fig:fede} and associated discussions for more details).} 
\label{fig:scalar}
\end{figure}

The initial conditions for the field, its velocity, and the Hubble expansion rate are chosen as discussed in section~\ref{sec_enh}.
Once these are set, we have to choose values for the parameters describing the potential~(\ref{secvP}).
Since the EDE energy density must decay faster than the energy density of matter or radiation, the value of $n$ is restricted to be $\geq 3$.
In our analysis, we fix this value to be $n=3$.
The other two parameters relevant for EDE, \ie $V_{0_{\rm ede}}$ and $f_1$, are chosen such that we satisfy the two conditions that the EDE component contributes $\sim 12\%$ of the total energy density of the universe, and the maximum of the EDE energy density occurs around the time of matter-radiation equality. 
The parameters $V_{0_{\rm de}}$ and $f_2$ are similarly chosen such that the onset of LDE domination occurs recently and at the correct energy scale so as to account for the current accelerated expansion of the universe.

The value
for the EDE decay constant $f_1$ turns out to be sub-Planckian, 
while we choose the value of the LDE decay constant $f_2$ as
\be\label{efs}
f_2 \simeq \frac{2 q }{2m+1} \,f_1\,,
\ee
with $q, m$ being integers. The quantity $q$ is determined by the periodicity of the EDE potential, $2\pi f_1$, and the  value of $\varphi$ is determined at an epoch where the DBI kinetic effects   end. By choosing a sufficiently large $m$, the value of $f_2$ relevant for LDE can be made sufficiently small in Planck units potentially satisfying the 
 constraints from the weak gravity conjecture for EDE \cite{Rudelius:2022gyu}.  For the values in Table  \ref{tab2} we have $q=6$ and we choose $m=20$. 

\begin{figure}
\begin{center}
\includegraphics[width=9.5cm]{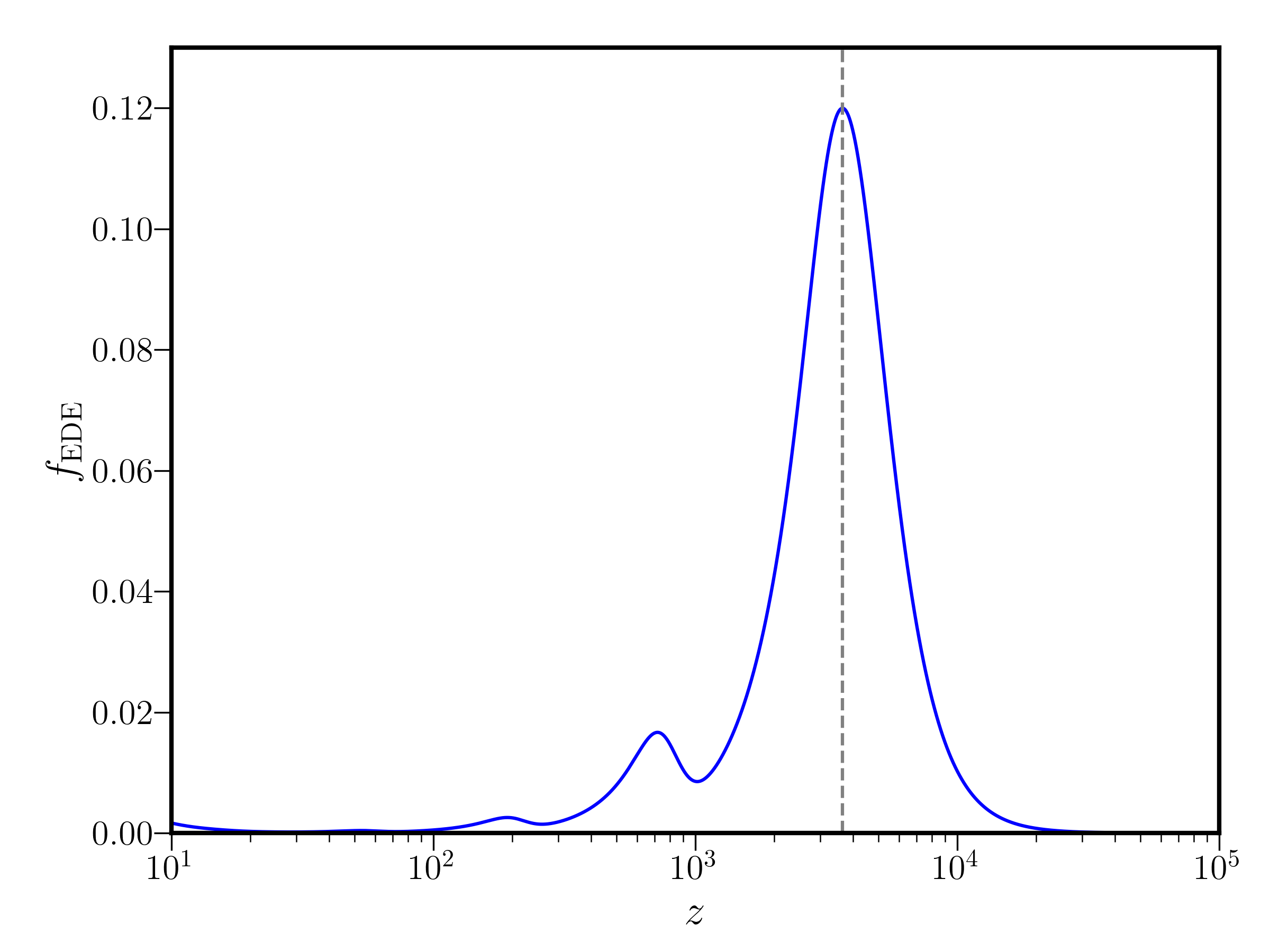}\\
\end{center}
\caption{\small The fractional contribution of the EDE to the total energy density has been plotted as a function of redshift. It reaches a maximum value of $\sim 0.12$ around $\log_{10} z\sim 3.562$, which has been shown to help in relaxing the Hubble tension~\cite{Hill:2020osr}.}
\label{fig:fede}
\end{figure}

Figure \ref{fig:scalar} shows
the evolution of the scalar field as a function of the universe temperature, starting from the initial conditions after inflation of Table~\ref{tab1}. During the initial epoch of DBI-kination, the field is unaffected by the potential. The axion potential becomes relevant at around  $T\sim $ eV, where the  field value remains  frozen by the Hubble friction acting as a cosmological constant: this corresponds to the phase
of EDE domination.  In order to satisfy  current constraints, and help in ameliorating
the Hubble tension,
the fractional energy density contributed by the EDE component,
\be\label{fede}
f_{\rm EDE}(z) \equiv \frac{\rho_{\rm EDE}}{\rho_T}\,,
\ee
 should not exceed  values around  $12\%$ (see e.g. \cite{Kamionkowski:2022pkx,Poulin:2023lkg}). This requirement is satisfied in our set-up, see Fig \ref{fig:fede}. It would nevertheless be important to analyse in more detail whether our modified
 evolution equation is in agreement with current
 CMB constraints: this goes beyond the scope of this work, and we defer it to a separate study.

\begin{figure}[h!]
\begin{center}
\includegraphics[width=9.5cm]{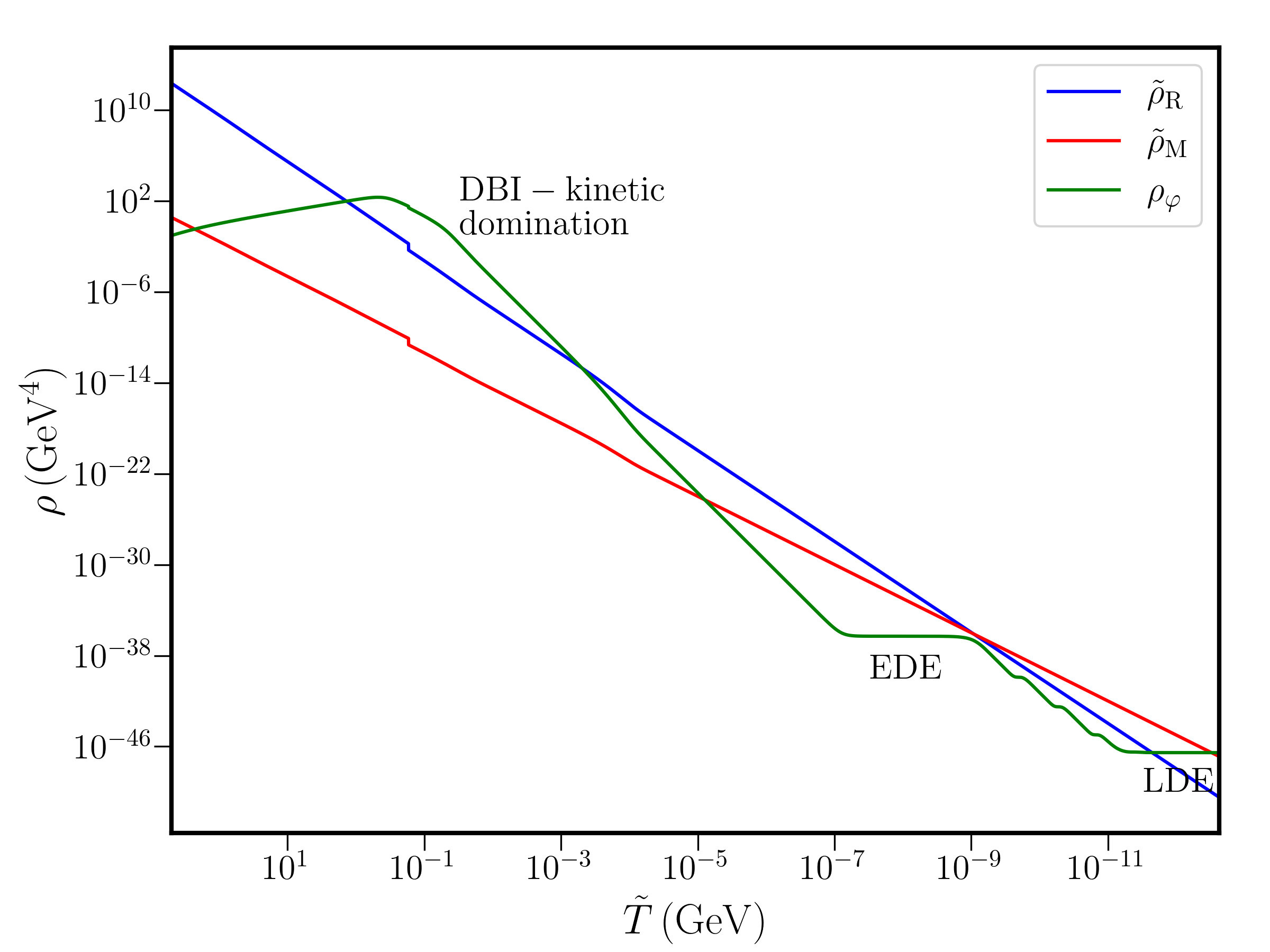}\\
\end{center}
\caption{\small Plot of the different contributions to the energy density of the universe  as a function  of the universe temperature. The  behaviour of the scalar field energy density at different temperatures  illustrates the DBI kination epoch -- the scalar energy density transiently dominates over the other energy densities  at high temperatures --  the
 EDE epoch -- the scalar energy density is constant and subdominant at intermediate temperatures --  and the LDE epoch --  the scalar energy density  returns to a constant value
 driving present-day acceleration.}
\label{fig:energyD}
\end{figure}

As the temperature of the universe decreases, the axion field rolls to the minimum of the EDE potential, around which it oscillates   until it feels the second LDE potential in eq \eqref{secvP}. This leads to the current cosmological acceleration. In Figure \ref{fig:energyD} we plot the evolution of the separate contributions to the total energy density of the universe from radiation, matter, and the axion field. The figure shows three  epochs driven by the D-brane axionic field:  DBI kination, EDE, and LDE. During the first phase of DBI kination the
scalar contribution briefly dominates the energy density; in the  phase of EDE the scalar
energy density is almost constant, and subdominant; in the last phase of LDE the scalar
energy density is again constant, and eventually starts to dominate the total energy density.

\section{Discussion}

The recent detection of a stochastic
GW background by several PTA collaborations raises
the question of its origin. In this work
we explored the possibility that it is sourced by  cosmic inflation, a well-known
source of primordial GW. The inflationary GW background
 amplitude is enhanced at PTA scales by a non-standard
 early cosmological evolution, driven by a kinetically-dominated scalar-tensor
 dynamics motivated by string theory.    The resulting 
 GW energy density has a distinctive broken power-law frequency profile,  entering
   the PTA band  with
  a peak amplitude consistent with the recent GW detections. Our framework --  besides
  providing a possible explanation for the recent PTA results -- also sheds light on
  other  well-known
  cosmological problems. It provides a realization of an early dark energy scenario aimed
  to address the $H_0$ tension, and a late dark energy model which explains
the   current cosmological acceleration with no need of a cosmological constant.

Several questions are left for further exploring this connection
between GW in the PTA band and the physics of the dark universe.
  It will be interesting to test in detail our
specific broken power-law frequency profile for the GW energy density against the actual PTA
data that are being currently released. It  would also be important 
to develop
 more careful  tests
of the early dark energy scenario we propose, for example comparing it against constraints from CMB observations, to further restrict our parameter space. This 
 analysis
 would require to modify existing CMB codes to accommodate the details of 
 our set-up. 
Since
 the characteristic  energy scale of the mechanism we propose is set around the scale 
 of QCD   transition, our scenario can have further consequences for the predictions
 of dark matter relic density  and sterile neutrinos (see e.g. \cite{Gelmini:2013awa}). We leave some of these investigations
 to a forthcoming work.

\subsection*{Acknowledgments}
We are
  partially funded by the STFC grant ST/T000813/1. For the purpose of open access, the authors have applied a Creative Commons Attribution licence to any Author Accepted Manuscript version arising.

\begin{appendix}

\section{Determining the initial condition for $H$}\label{App1}

In this appendix we briefly describe how we fix the initial conditions for the Hubble
parameter in our analysis.
We define the Hubble parameter in the standard $\Lambda$CDM scenario as
\begin{equation}
 H_{\Lambda {\rm CDM}}^2 = \frac{\kappa^2}{3}\,\left(\tilde{\rho}_{\rm R}+\tilde{\rho}_{\rm M}+\tilde{\rho}_{\Lambda}\right) = \frac{\kappa^2}{3}\,\tilde{\rho}_{\rm bg}.
 \end{equation}
 The Hubble parameter in the disformal DBI case can be expressed as 
 \begin{subequations}
\begin{align}
 H^2 &= \frac{\kappa^2}{3}\,\left(\rho_{_{\varphi}}+\rho_{\rm bg}\right) \nonumber\\
 &= \frac{\kappa^2}{3\,B}\,\left(\,\gamma\,\tilde{\rho_{\rm bg}} + V\right).
 \end{align}
\end{subequations}
 Using the expression~(\ref{eq:B-Dbrane}), we can write
\begin{equation}
 H^2\left[1-\frac{\gamma^2\,\varphi_{_N}^2}{3\,(\gamma+1)}\right] = \frac{\kappa^2}{3}\,\left(\,\gamma\,\tilde{\rho_{\rm bg}} + V\right).
\end{equation}
Substituting the relation~(\ref{eq:gamma-Dbrane}),
we get the following quartic equation in $\gamma$:
\begin{eqnarray}
 \gamma^4\left(M^{-4}\,\varphi_{_N}^2\,\,\tilde{\rho}_{\rm bg} + \varphi_{_N}^2\right)&+&\gamma^3\left(M^{-4}\,\varphi_{_N}^2\,V-3\,\right)+\gamma^2\left(M^{-4}\,\varphi_{_N}^2\,\,\tilde{\rho}_{\rm bg}-3\,-\,\varphi_{_N}^2\right) \nonumber\\
 &+&\gamma\left(M^{-4}\,\varphi_{_N}^2\,V+3\,\right)+3\,=0.
\end{eqnarray}
This equation can be solved to obtain four initial values of $\gamma$ for a given set of initial conditions. For our system, we choose the value of $\gamma_i$ to be real and of
order one.  With this choice of $\gamma_i$, we can then obtain the value of $H_i$ from the relation~(\ref{eq:gamma-Dbrane}).

\end{appendix}

\bibliographystyle{JHEP}
\bibliography{nano}

\providecommand{\href}[2]{#2}\begingroup\raggedright\begin{thebibliography}{10}

\bibitem{Grishchuk:1974ny}
L.~P. Grishchuk, {\it {Amplification of gravitational waves in an istropic
  universe}},  {\em Zh. Eksp. Teor. Fiz.} {\bf 67} (1974) 825--838.

\bibitem{Starobinsky:1979ty}
A.~A. Starobinsky, {\it {Spectrum of relict gravitational radiation and the
  early state of the universe}},  {\em JETP Lett.} {\bf 30} (1979) 682--685.

\bibitem{Rubakov:1982df}
V.~A. Rubakov, M.~V. Sazhin, and A.~V. Veryaskin, {\it {Graviton Creation in
  the Inflationary Universe and the Grand Unification Scale}},  {\em Phys.
  Lett. B} {\bf 115} (1982) 189--192.

\bibitem{Fabbri:1983us}
R.~Fabbri and M.~d. Pollock, {\it {The Effect of Primordially Produced
  Gravitons upon the Anisotropy of the Cosmological Microwave Background
  Radiation}},  {\em Phys. Lett. B} {\bf 125} (1983) 445--448.

\bibitem{Abbott:1984fp}
L.~F. Abbott and M.~B. Wise, {\it {Constraints on Generalized Inflationary
  Cosmologies}},  {\em Nucl. Phys. B} {\bf 244} (1984) 541--548.

\bibitem{Haque:2021dha}
M.~R. Haque, D.~Maity, T.~Paul, and L.~Sriramkumar, {\it {Decoding the phases
  of early and late time reheating through imprints on primordial gravitational
  waves}},  {\em Phys. Rev. D} {\bf 104} (2021), no.~6 063513,
  [\href{http://arxiv.org/abs/2105.09242}{{\tt arXiv:2105.09242}}].

\bibitem{Ashoorioon:2022raz}
A.~Ashoorioon, K.~Rezazadeh, and A.~Rostami, {\it {NANOGrav signal from the end
  of inflation and the LIGO mass and heavier primordial black holes}},  {\em
  Phys. Lett. B} {\bf 835} (2022) 137542,
  [\href{http://arxiv.org/abs/2202.01131}{{\tt arXiv:2202.01131}}].

\bibitem{Guo:2023hyp}
S.-Y. Guo, M.~Khlopov, X.~Liu, L.~Wu, Y.~Wu, and B.~Zhu, {\it {Footprints of
  Axion-Like Particle in Pulsar Timing Array Data and JWST Observations}},
  \href{http://arxiv.org/abs/2306.17022}{{\tt arXiv:2306.17022}}.

\bibitem{Giovannini:1998bp}
M.~Giovannini, {\it {Gravitational waves constraints on postinflationary phases
  stiffer than radiation}},  {\em Phys. Rev. D} {\bf 58} (1998) 083504,
  [\href{http://arxiv.org/abs/hep-ph/9806329}{{\tt hep-ph/9806329}}].

\bibitem{Boyle:2007zx}
L.~A. Boyle and A.~Buonanno, {\it {Relating gravitational wave constraints from
  primordial nucleosynthesis, pulsar timing, laser interferometers, and the
  CMB: Implications for the early Universe}},  {\em Phys. Rev. D} {\bf 78}
  (2008) 043531, [\href{http://arxiv.org/abs/0708.2279}{{\tt
  arXiv:0708.2279}}].

\bibitem{Boyle:2005se}
L.~A. Boyle and P.~J. Steinhardt, {\it {Probing the early universe with
  inflationary gravitational waves}},  {\em Phys. Rev. D} {\bf 77} (2008)
  063504, [\href{http://arxiv.org/abs/astro-ph/0512014}{{\tt
  astro-ph/0512014}}].

\bibitem{Salati:2002md}
P.~Salati, {\it {Quintessence and the relic density of neutralinos}},  {\em
  Phys. Lett. B} {\bf 571} (2003) 121--131,
  [\href{http://arxiv.org/abs/astro-ph/0207396}{{\tt astro-ph/0207396}}].

\bibitem{Catena:2004ba}
R.~Catena, N.~Fornengo, A.~Masiero, M.~Pietroni, and F.~Rosati, {\it {Dark
  matter relic abundance and scalar - tensor dark energy}},  {\em Phys. Rev. D}
  {\bf 70} (2004) 063519, [\href{http://arxiv.org/abs/astro-ph/0403614}{{\tt
  astro-ph/0403614}}].

\bibitem{Gelmini:2013awa}
G.~B. Gelmini, J.-H. Huh, and T.~Rehagen, {\it {Asymmetric dark matter
  annihilation as a test of non-standard cosmologies}},  {\em JCAP} {\bf 08}
  (2013) 003, [\href{http://arxiv.org/abs/1304.3679}{{\tt arXiv:1304.3679}}].

\bibitem{Lahanas:2006hf}
A.~B. Lahanas, N.~E. Mavromatos, and D.~V. Nanopoulos, {\it {Dilaton and
  off-shell (non-critical string) effects in Boltzmann equation for species
  abundances}},  {\em PMC Phys. A} {\bf 1} (2007) 2,
  [\href{http://arxiv.org/abs/hep-ph/0608153}{{\tt hep-ph/0608153}}].

\bibitem{Pallis:2009ed}
C.~Pallis, {\it {Cold Dark Matter in non-Standard Cosmologies, PAMELA, ATIC and
  Fermi LAT}},  {\em Nucl. Phys. B} {\bf 831} (2010) 217--247,
  [\href{http://arxiv.org/abs/0909.3026}{{\tt arXiv:0909.3026}}].

\bibitem{Dutta:2016htz}
B.~Dutta, E.~Jimenez, and I.~Zavala, {\it {Dark Matter Relics and the Expansion
  Rate in Scalar-Tensor Theories}},  {\em JCAP} {\bf 06} (2017) 032,
  [\href{http://arxiv.org/abs/1612.05553}{{\tt arXiv:1612.05553}}].

\bibitem{Dutta:2017fcn}
B.~Dutta, E.~Jimenez, and I.~Zavala, {\it {D-brane Disformal Coupling and
  Thermal Dark Matter}},  {\em Phys. Rev. D} {\bf 96} (2017), no.~10 103506,
  [\href{http://arxiv.org/abs/1708.07153}{{\tt arXiv:1708.07153}}].

\bibitem{Chowdhury:2022gdc}
D.~Chowdhury, G.~Tasinato, and I.~Zavala, {\it {The rise of the primordial
  tensor spectrum from an early scalar-tensor epoch}},  {\em JCAP} {\bf 08}
  (2022), no.~08 010, [\href{http://arxiv.org/abs/2204.10218}{{\tt
  arXiv:2204.10218}}].

\bibitem{NANOGrav:2023gor}
{\bf NANOGrav} Collaboration, G.~Agazie et~al., {\it {The NANOGrav 15-year Data
  Set: Evidence for a Gravitational-Wave Background}},  {\em Astrophys. J.
  Lett.} {\bf 951} (2023), no.~1 [\href{http://arxiv.org/abs/2306.16213}{{\tt
  arXiv:2306.16213}}].

\bibitem{Reardon:2023gzh}
D.~J. Reardon et~al., {\it {Search for an isotropic gravitational-wave
  background with the Parkes Pulsar Timing Array}},  {\em Astrophys. J. Lett.}
  {\bf 951} (2023), no.~1 [\href{http://arxiv.org/abs/2306.16215}{{\tt
  arXiv:2306.16215}}].

\bibitem{Antoniadis:2023ott}
J.~Antoniadis et~al., {\it {The second data release from the European Pulsar
  Timing Array III. Search for gravitational wave signals}},
  \href{http://arxiv.org/abs/2306.16214}{{\tt arXiv:2306.16214}}.

\bibitem{Xu:2023wog}
H.~Xu et~al., {\it {Searching for the Nano-Hertz Stochastic Gravitational Wave
  Background with the Chinese Pulsar Timing Array Data Release I}},  {\em Res.
  Astron. Astrophys.} {\bf 23} (2023), no.~7 075024,
  [\href{http://arxiv.org/abs/2306.16216}{{\tt arXiv:2306.16216}}].

\bibitem{NANOGrav:2023hvm}
{\bf NANOGrav} Collaboration, A.~Afzal et~al., {\it {The NANOGrav 15-year Data
  Set: Search for Signals from New Physics}},  {\em Astrophys. J. Lett.} {\bf
  951} (2023), no.~1 [\href{http://arxiv.org/abs/2306.16219}{{\tt
  arXiv:2306.16219}}].

\bibitem{Antoniadis:2023xlr}
J.~Antoniadis et~al., {\it {The second data release from the European Pulsar
  Timing Array: V. Implications for massive black holes, dark matter and the
  early Universe}},  \href{http://arxiv.org/abs/2306.16227}{{\tt
  arXiv:2306.16227}}.

\bibitem{DiValentino:2021izs}
E.~Di~Valentino, O.~Mena, S.~Pan, L.~Visinelli, W.~Yang, A.~Melchiorri, D.~F.
  Mota, A.~G. Riess, and J.~Silk, {\it {In the realm of the Hubble
  tension\textemdash{}a review of solutions}},  {\em Class. Quant. Grav.} {\bf
  38} (2021), no.~15 153001, [\href{http://arxiv.org/abs/2103.01183}{{\tt
  arXiv:2103.01183}}].

\bibitem{Schoneberg:2021qvd}
N.~Sch\"oneberg, G.~Franco~Abell\'an, A.~P\'erez~S\'anchez, S.~J. Witte,
  V.~Poulin, and J.~Lesgourgues, {\it {The H0 Olympics: A fair ranking of
  proposed models}},  {\em Phys. Rept.} {\bf 984} (2022) 1--55,
  [\href{http://arxiv.org/abs/2107.10291}{{\tt arXiv:2107.10291}}].

\bibitem{Bekenstein:1992pj}
J.~D. Bekenstein, {\it {The Relation between physical and gravitational
  geometry}},  {\em Phys. Rev. D} {\bf 48} (1993) 3641--3647,
  [\href{http://arxiv.org/abs/gr-qc/9211017}{{\tt gr-qc/9211017}}].

\bibitem{Rezazadeh:2022lsf}
K.~Rezazadeh, A.~Ashoorioon, and D.~Grin, {\it {Cascading Dark Energy}},
  \href{http://arxiv.org/abs/2208.07631}{{\tt arXiv:2208.07631}}.

\bibitem{Cicoli:2023opf}
M.~Cicoli, J.~P. Conlon, A.~Maharana, S.~Parameswaran, F.~Quevedo, and
  I.~Zavala, {\it {String Cosmology: from the Early Universe to Today}},
  \href{http://arxiv.org/abs/2303.04819}{{\tt arXiv:2303.04819}}.

\bibitem{Koivisto:2013fta}
T.~Koivisto, D.~Wills, and I.~Zavala, {\it {Dark D-brane Cosmology}},  {\em
  JCAP} {\bf 06} (2014) 036, [\href{http://arxiv.org/abs/1312.2597}{{\tt
  arXiv:1312.2597}}].

\bibitem{Sakstein:2014isa}
J.~Sakstein, {\it {Disformal Theories of Gravity: From the Solar System to
  Cosmology}},  {\em JCAP} {\bf 12} (2014) 012,
  [\href{http://arxiv.org/abs/1409.1734}{{\tt arXiv:1409.1734}}].

\bibitem{vandeBruck:2020fjo}
C.~van~de Bruck and E.~M. Teixeira, {\it {Dark D-Brane Cosmology: from
  background evolution to cosmological perturbations}},  {\em Phys. Rev. D}
  {\bf 102} (2020), no.~10 103503, [\href{http://arxiv.org/abs/2007.15414}{{\tt
  arXiv:2007.15414}}].

\bibitem{Polchinski:1996na}
J.~Polchinski, {\it {Tasi lectures on D-branes}},  in {\em {Theoretical
  Advanced Study Institute in Elementary Particle Physics (TASI 96): Fields,
  Strings, and Duality}}, pp.~293--356, 11, 1996.
\newblock \href{http://arxiv.org/abs/hep-th/9611050}{{\tt hep-th/9611050}}.

\bibitem{Kenton:2014gma}
Z.~Kenton and S.~Thomas, {\it {D-brane Potentials in the Warped Resolved
  Conifold and Natural Inflation}},  {\em JHEP} {\bf 02} (2015) 127,
  [\href{http://arxiv.org/abs/1409.1221}{{\tt arXiv:1409.1221}}].

\bibitem{Chakraborty:2019dfh}
D.~Chakraborty, R.~Chiovoloni, O.~Loaiza-Brito, G.~Niz, and I.~Zavala, {\it
  {Fat inflatons, large turns and the $\eta$-problem}},  {\em JCAP} {\bf 01}
  (2020) 020, [\href{http://arxiv.org/abs/1908.09797}{{\tt arXiv:1908.09797}}].

\bibitem{Candel}
P.~Candelas and X.~C. de~la Ossa, {\it {Comments on Conifolds}},  {\em Nucl.
  Phys.} {\bf B342} (1990) 246--268.

\bibitem{RCmetric}
L.~A. Pando~Zayas and A.~A. Tseytlin, {\it {3-branes on resolved conifold}},
  {\em JHEP} {\bf 11} (2000) 028,
  [\href{http://arxiv.org/abs/hep-th/0010088}{{\tt hep-th/0010088}}].

\bibitem{Kleb}
I.~R. Klebanov and A.~Murugan, {\it {Gauge/Gravity Duality and Warped Resolved
  Conifold}},  {\em JHEP} {\bf 03} (2007) 042,
  [\href{http://arxiv.org/abs/hep-th/0701064}{{\tt hep-th/0701064}}].

\bibitem{Baumann:2010sx}
D.~Baumann, A.~Dymarsky, S.~Kachru, I.~R. Klebanov, and L.~McAllister, {\it
  {D3-brane Potentials from Fluxes in AdS/CFT}},  {\em JHEP} {\bf 06} (2010)
  072, [\href{http://arxiv.org/abs/1001.5028}{{\tt arXiv:1001.5028}}].

\bibitem{Poulin:2018dzj}
V.~Poulin, T.~L. Smith, D.~Grin, T.~Karwal, and M.~Kamionkowski, {\it
  {Cosmological implications of ultralight axionlike fields}},  {\em Phys. Rev.
  D} {\bf 98} (2018), no.~8 083525,
  [\href{http://arxiv.org/abs/1806.10608}{{\tt arXiv:1806.10608}}].

\bibitem{Poulin:2018cxd}
V.~Poulin, T.~L. Smith, T.~Karwal, and M.~Kamionkowski, {\it {Early Dark Energy
  Can Resolve The Hubble Tension}},  {\em Phys. Rev. Lett.} {\bf 122} (2019),
  no.~22 221301, [\href{http://arxiv.org/abs/1811.04083}{{\tt
  arXiv:1811.04083}}].

\bibitem{Kaloper:2005aj}
N.~Kaloper and L.~Sorbo, {\it {Of pngb quintessence}},  {\em JCAP} {\bf 04}
  (2006) 007, [\href{http://arxiv.org/abs/astro-ph/0511543}{{\tt
  astro-ph/0511543}}].

\bibitem{Planck:2015bue}
{\bf Planck} Collaboration, P.~A.~R. Ade et~al., {\it {Planck 2015 results.
  XIV. Dark energy and modified gravity}},  {\em Astron. Astrophys.} {\bf 594}
  (2016) A14, [\href{http://arxiv.org/abs/1502.01590}{{\tt arXiv:1502.01590}}].

\bibitem{Damour:1993id}
T.~Damour and K.~Nordtvedt, {\it {Tensor - scalar cosmological models and their
  relaxation toward general relativity}},  {\em Phys. Rev. D} {\bf 48} (1993)
  3436--3450.

\bibitem{Ferreira:1997hj}
P.~G. Ferreira and M.~Joyce, {\it {Cosmology with a primordial scaling field}},
   {\em Phys. Rev. D} {\bf 58} (1998) 023503,
  [\href{http://arxiv.org/abs/astro-ph/9711102}{{\tt astro-ph/9711102}}].

\bibitem{Redmond:2018xty}
K.~Redmond, A.~Trezza, and A.~L. Erickcek, {\it {Growth of Dark Matter
  Perturbations during Kination}},  {\em Phys. Rev. D} {\bf 98} (2018), no.~6
  063504, [\href{http://arxiv.org/abs/1807.01327}{{\tt arXiv:1807.01327}}].

\bibitem{Co:2021lkc}
R.~T. Co, D.~Dunsky, N.~Fernandez, A.~Ghalsasi, L.~J. Hall, K.~Harigaya, and
  J.~Shelton, {\it {Gravitational wave and CMB probes of axion kination}},
  {\em JHEP} {\bf 09} (2022) 116, [\href{http://arxiv.org/abs/2108.09299}{{\tt
  arXiv:2108.09299}}].

\bibitem{Gouttenoire:2021jhk}
Y.~Gouttenoire, G.~Servant, and P.~Simakachorn, {\it {Kination cosmology from
  scalar fields and gravitational-wave signatures}},
  \href{http://arxiv.org/abs/2111.01150}{{\tt arXiv:2111.01150}}.

\bibitem{Bernal:2020ywq}
N.~Bernal, A.~Ghoshal, F.~Hajkarim, and G.~Lambiase, {\it {Primordial
  Gravitational Wave Signals in Modified Cosmologies}},  {\em JCAP} {\bf 11}
  (2020) 051, [\href{http://arxiv.org/abs/2008.04959}{{\tt arXiv:2008.04959}}].

\bibitem{Watanabe:2006qe}
Y.~Watanabe and E.~Komatsu, {\it {Improved Calculation of the Primordial
  Gravitational Wave Spectrum in the Standard Model}},  {\em Phys. Rev. D} {\bf
  73} (2006) 123515, [\href{http://arxiv.org/abs/astro-ph/0604176}{{\tt
  astro-ph/0604176}}].

\bibitem{Saikawa:2018rcs}
K.~Saikawa and S.~Shirai, {\it {Primordial gravitational waves, precisely: The
  role of thermodynamics in the Standard Model}},  {\em JCAP} {\bf 05} (2018)
  035, [\href{http://arxiv.org/abs/1803.01038}{{\tt arXiv:1803.01038}}].

\bibitem{Bernal:2019lpc}
N.~Bernal and F.~Hajkarim, {\it {Primordial Gravitational Waves in Nonstandard
  Cosmologies}},  {\em Phys. Rev. D} {\bf 100} (2019), no.~6 063502,
  [\href{http://arxiv.org/abs/1905.10410}{{\tt arXiv:1905.10410}}].

\bibitem{BICEP:2021xfz}
{\bf BICEP, Keck} Collaboration, P.~A.~R. Ade et~al., {\it {Improved
  Constraints on Primordial Gravitational Waves using Planck, WMAP, and
  BICEP/Keck Observations through the 2018 Observing Season}},  {\em Phys. Rev.
  Lett.} {\bf 127} (2021), no.~15 151301,
  [\href{http://arxiv.org/abs/2110.00483}{{\tt arXiv:2110.00483}}].

\bibitem{Kamionkowski:1993fg}
M.~Kamionkowski, A.~Kosowsky, and M.~S. Turner, {\it {Gravitational radiation
  from first order phase transitions}},  {\em Phys. Rev. D} {\bf 49} (1994)
  2837--2851, [\href{http://arxiv.org/abs/astro-ph/9310044}{{\tt
  astro-ph/9310044}}].

\bibitem{NANOGrav:2023ctt}
{\bf NANOGrav} Collaboration, G.~Agazie et~al., {\it {The NANOGrav 15 yr Data
  Set: Detector Characterization and Noise Budget}},  {\em Astrophys. J. Lett.}
  {\bf 951} (2023), no.~1 L10, [\href{http://arxiv.org/abs/2306.16218}{{\tt
  arXiv:2306.16218}}].

\bibitem{the_nanograv_collaboration_2023_8092346}
{The NANOGrav Collaboration}, {\it {Noise Spectra and Stochastic Background
  Sensitivity Curve for the NG15-year Dataset}}, .

\bibitem{Thrane:2013oya}
E.~Thrane and J.~D. Romano, {\it {Sensitivity curves for searches for
  gravitational-wave backgrounds}},  {\em Phys. Rev. D} {\bf 88} (2013), no.~12
  124032, [\href{http://arxiv.org/abs/1310.5300}{{\tt arXiv:1310.5300}}].

\bibitem{Schmitz:2020syl}
K.~Schmitz, {\it {New Sensitivity Curves for Gravitational-Wave Signals from
  Cosmological Phase Transitions}},  {\em JHEP} {\bf 01} (2021) 097,
  [\href{http://arxiv.org/abs/2002.04615}{{\tt arXiv:2002.04615}}].

\bibitem{NANOGrav:2020bcs}
{\bf NANOGrav} Collaboration, Z.~Arzoumanian et~al., {\it {The NANOGrav 12.5 yr
  Data Set: Search for an Isotropic Stochastic Gravitational-wave Background}},
   {\em Astrophys. J. Lett.} {\bf 905} (2020), no.~2 L34,
  [\href{http://arxiv.org/abs/2009.04496}{{\tt arXiv:2009.04496}}].

\bibitem{andrea_mitridate_2023_8106173}
A.~Mitridate and D.~Wright, {\it {PTArcade}}, .

\bibitem{Mitridate:2023oar}
A.~Mitridate, D.~Wright, R.~von Eckardstein, T.~Schr\"oder, J.~Nay, K.~Olum,
  K.~Schmitz, and T.~Trickle, {\it {PTArcade}},
  \href{http://arxiv.org/abs/2306.16377}{{\tt arXiv:2306.16377}}.

\bibitem{Lamb:2023jls}
W.~G. Lamb, S.~R. Taylor, and R.~van Haasteren, {\it {The Need For Speed: Rapid
  Refitting Techniques for Bayesian Spectral Characterization of the
  Gravitational Wave Background Using PTAs}},
  \href{http://arxiv.org/abs/2303.15442}{{\tt arXiv:2303.15442}}.

\bibitem{Kitajima:2023mxn}
N.~Kitajima and T.~Takahashi, {\it {Stochastic gravitational wave background
  from early dark energy}},  \href{http://arxiv.org/abs/2306.16896}{{\tt
  arXiv:2306.16896}}.

\bibitem{Wetterich:2004pv}
C.~Wetterich, {\it {Phenomenological parameterization of quintessence}},  {\em
  Phys. Lett. B} {\bf 594} (2004) 17--22,
  [\href{http://arxiv.org/abs/astro-ph/0403289}{{\tt astro-ph/0403289}}].

\bibitem{Doran:2006kp}
M.~Doran and G.~Robbers, {\it {Early dark energy cosmologies}},  {\em JCAP}
  {\bf 06} (2006) 026, [\href{http://arxiv.org/abs/astro-ph/0601544}{{\tt
  astro-ph/0601544}}].

\bibitem{Kamionkowski:2022pkx}
M.~Kamionkowski and A.~G. Riess, {\it {The Hubble Tension and Early Dark
  Energy}},  \href{http://arxiv.org/abs/2211.04492}{{\tt arXiv:2211.04492}}.

\bibitem{Poulin:2023lkg}
V.~Poulin, T.~L. Smith, and T.~Karwal, {\it {The Ups and Downs of Early Dark
  Energy solutions to the Hubble tension: a review of models, hints and
  constraints circa 2023}},  \href{http://arxiv.org/abs/2302.09032}{{\tt
  arXiv:2302.09032}}.

\bibitem{Kim:2002tq}
J.~E. Kim and H.~P. Nilles, {\it {A Quintessential axion}},  {\em Phys. Lett.
  B} {\bf 553} (2003) 1--6, [\href{http://arxiv.org/abs/hep-ph/0210402}{{\tt
  hep-ph/0210402}}].

\bibitem{McDonough:2022pku}
E.~McDonough and M.~Scalisi, {\it {Towards Early Dark Energy in String
  Theory}},  \href{http://arxiv.org/abs/2209.00011}{{\tt arXiv:2209.00011}}.

\bibitem{Cicoli:2023qri}
M.~Cicoli, M.~Licheri, R.~Mahanta, E.~McDonough, F.~G. Pedro, and M.~Scalisi,
  {\it {Early Dark Energy in Type IIB String Theory}},  {\em JHEP} {\bf 06}
  (2023) 052, [\href{http://arxiv.org/abs/2303.03414}{{\tt arXiv:2303.03414}}].

\bibitem{Rudelius:2022gyu}
T.~Rudelius, {\it {Constraints on early dark energy from the axion weak gravity
  conjecture}},  {\em JCAP} {\bf 01} (2023) 014,
  [\href{http://arxiv.org/abs/2203.05575}{{\tt arXiv:2203.05575}}].

\bibitem{Hill:2020osr}
J.~C. Hill, E.~McDonough, M.~W. Toomey, and S.~Alexander, {\it {Early dark
  energy does not restore cosmological concordance}},  {\em Phys. Rev. D} {\bf
  102} (2020), no.~4 043507, [\href{http://arxiv.org/abs/2003.07355}{{\tt
  arXiv:2003.07355}}].

\end{thebibliography}\endgroup
\end{document}